\documentclass[showpacs,aps,fleqn,onecolumn]{revtex4}%
\usepackage{graphicx}
\usepackage{psfrag}
\usepackage{amsmath}

\begin{document}
\def\be{\begin{eqnarray}}
\def\ee{\end{eqnarray}}
\newcommand{\nn}{\nonumber}
\def\mpcomm#1{\nextline\strut\kern-6em{\tt MP COMMENT => \ #1}\nextline}
\def\nextline{\hfill\break}
\newcommand{\rf}[1]{(\ref{#1})}
\newcommand{\beq}{\begin{equation}}
\newcommand{\eeq}{\end{equation}}
\newcommand{\bea}{\begin{eqnarray}}
\newcommand{\eea}{\end{eqnarray}}
\newcommand{\pint}{-\hspace{-10pt}\int_{-\infty}^\infty }
\newcommand{\Pint}{-\hspace{-11pt}\int_{-\infty}^\infty }
\newcommand{\nint}{\int_{\infty}^{\infty}}

\title{Random L\'{e}vy Matrices Revisited}

\author{Zdzis\l{}aw Burda$^{1,2}$}{\thanks{burda@th.if.uj.edu.pl}}
\author{Jerzy Jurkiewicz$^{1,2}$}{\thanks{jjurkiew@th.if.uj.edu.pl}}
\author{Maciej A. Nowak$^{1,2}$}{\thanks{nowak@th.if.uj.edu.pl}}
\author{Gabor Papp$^{3}$}{\thanks{pg@ludens.elte.hu}}
\author{Ismail Zahed$^{4}$}{\thanks{zahed@zahed.physics.sunysb.edu}}

\affiliation{ \bigskip
$^1$Marian Smoluchowski Institute of Physics, 
Jagiellonian University,
30-059 Krak\'{o}w, Reymonta 4, Poland \\
$^2$Mark Kac Complex Systems Research Centre,
Jagiellonian University, Krak\'{o}w, Poland \\
$^3$Institute of Physics, E\"{o}tv\"{o}s University,
P\'{a}zm\'{a}ny P.s.1/a, H-1117 Budapest, Hungary\\
$^4$Department of Physics and Astronomy, SUNY Stony Brook, NY
11794, USA }
{\begin{abstract}
We compare eigenvalue densities of Wigner random matrices whose
elements are independent identically distributed (iid) random
numbers with a L\'evy distribution and maximally
random matrices with a rotationally invariant
measure exhibiting a power law spectrum given by stable laws
of free random variables. We compute the eigenvalue density
of Wigner-L\'evy (WL) matrices using (and correcting)
the method by Bouchaud and Cizeau (BC), and of free random L\'evy (FRL) 
rotationally invariant matrices by adapting
results of free probability calculus. 
We compare the two types of eigenvalue spectra.
Both ensembles are spectrally stable with respect 
to the matrix addition. The discussed ensemble of FRL matrices
is maximally random in the sense that it maximizes Shannon's entropy.
We find a perfect agreement between the numerically sampled
spectra and the analytical results already for matrices
of dimension $N=100$. The numerical spectra show very weak dependence on
the matrix size $N$ as can be noticed by comparing spectra
for $N=400$. After a pertinent rescaling spectra of Wigner-L\'evy
matrices and of symmetric FRL matrices have the same tail behavior. 
As we discuss towards the end of the paper  
the correlations of large eigenvalues in the two ensembles are however
different. We illustrate the relation between the two types 
of stability and show that the addition of many randomly
rotated Wigner-L\'evy matrices leads by a matrix central
limit theorem to FRL spectra, providing an explicit realization
of the maximal randomness principle.

\end{abstract}
}

 \maketitle

\section{Introduction}

Applications of random matrix theory cover many
branches of physics and cross-disciplinary
fields~\cite{RMTGENERAL} involving multivariate analysis of large
and noisy data sets~\cite{MULTIVARIATE}. The standard random matrix
formulation belongs to the Gaussian basin, with a measure that is
Gaussian or polynomial with finite second moment. The ensuing
macroscopic spectral distribution is localized with finite supports
on the real axis. The canonical distribution for a Gaussian measure
is Wigner's semi-circle.

The class of stable (L\'{e}vy) distributions~\cite{LEVY} is however much
larger (the Gaussian class represents only one fixed point in the
stability basin of the L\'{e}vy class), and one is tempted to ask, why
the theory of random L\'{e}vy matrices is not so well established.
The case of L\'{e}vy randomness is far from being academic, and many
distributions in physics and outside (finance, networks) exhibit
power-like behavior referred to as {\it fat} or
{\it heavy tails}~\cite{FATTAIL}.

One of the chief reasons for why the theory of
random L\'{e}vy matrices is not yet well understood is the
technical difficulty inherent to these distributions. First,
even for one-dimensional stable distributions,
the explicit form of the probability distribution
functions (pdf) is known analytically only in
few cases~\cite{FELLER,ZOLOTARIEV}.
Second, L\'{e}vy distributions have divergent moments, and the
finiteness of the second moment (condition for the Gaussian stability
class) is usually a key for many of the techniques established in
Random Matrix Theory. Third, numerical studies involving
power-like behavior on infinite supports require enormous statistics
and are very sensitive to systematic errors.

In 1994, Bouchaud and Cizeau~\cite{BC} (hereafter BC)
considered large $N\times N$  random symmetric
matrices with entries sampled from one-dimensional, stable
distributions. In the large $N$ limit they obtained analytical
equations for the entries of the resolvent, and then
checked their predictions for the spectra using numerically
generated spectra by random sampling. The agreement was fair, although not
perfect. Contrary to the standard Gaussian-like ensembles, the
measure in the BC approach was not rotationally invariant.

In 2002, following the work in~\cite{BERVOIC}, we have suggested
another L\'{e}vy-type ensemble~\cite{FREELEVY} (hereafter FRL).
By construction its measure is rotationally invariant.
The average spectral distribution in this ensemble is stable
under matrix the convolution of two independent but identical
ensembles. It is similar to the stability property of 
one-dimensional L\'{e}vy distributions. 
The measure is non-analytic in the matrix $H$ and
universal at large $H$ with a potential $V(H)\approx {\rm ln}\,H^2$. This
weak logarithmic rise in the asymptotic potential is at the origin
of the long tail in the eigenvalue spectra. 

The present work will compare the WL and FRL results as advertised
in~\cite{KRACOW}. In section~2 we reanalyze and correct the original
arguments for the resolvent presented in~\cite{BC}. Our integral
equations for the resolvent and spectral density are different from
the ones in~\cite{BC}. We carry explicit analytical
transformations and expansions to provide insights to the spectrum.
We show that there is a perfect agreement between the analytical and
numerical results obtained by sampling large L\'{e}vy matrices. 
We also discuss the relation of ours and BC's results. 
In section~3 we recall the key concepts behind FRL ensembles. In large
$N$ the resolvent obeys a simple analytic equation. The resulting
spectra are compared to the spectra following from the corrected BC
analysis. 
The WL and FRL matrices represent two types of stability under
matrix convolution. In both cases we have a power behavior in the
tails of the spectrum. By a pertinent rescaling we may in fact
enforce the same tail behavior and compare the spectra. The observed
differences disappear in the Gauss limit and become more pronounced
in the Cauchy limit. In sectin 4 we explain the relation between 
the two types of stability on a simple example: we construct sums 
of WL matrices rotated by random $O(N)$ matrices and show that 
the spectrum of these sums converges by a matrix central 
limit theorem to the pertinent symmetric FRL spectrum. Our 
conclusions are in section 5.

\section{Wigner-L\'{e}vy Matrices}

\subsection*{Definition of the ensemble}
In a pioneering study on random L\'{e}vy matrices, Bouchaud and
Cizeau~\cite{BC} discussed a Wigner ensemble of $N\times N$
real symmetric random matrices with elements being iid random variables:
with probability density function following a L\'{e}vy
distribution $P(x)\equiv N^{1/\mu} L_\mu^{C,\beta}\left(N^{1/\mu} x\right)$,
with $\mu$ being the stability index, $\beta$ -- the asymmetry parameter,
and $C$ -- the range of the distribution (see below). We shall
call these matrices Wigner-L\'evy (WL) or Bouchaud-Cizeau (BC) matrices.
The probability measure for the ensemble of such matrices 
is given by:
\be
d\mu_{WL}(H) =\prod_{i\le j}\,P(H_{ij})\,dH_{ij}
\label{measure}
\ee
The scaling factor $N^{1/\mu}$ in pdf makes the
limiting eigenvalue density independent of the matrix 
size $N$ when $N\rightarrow \infty$. 
Alternatively one can think of the matrix elements
$H_{ij}$ as if they were calculated as 
$H_{ij} = h_{ij}/N^{1/\mu}$ with $h_{ij}$ being 
iid random numbers independent of $N$:
$p(x) = L_\mu^{C,\beta}(x)$.

L\'{e}vy distributions are notoriously hard to write explicitly
(except in few cases), but their characteristic functions are more
user friendly~\cite{FELLER}
\be
L_{\mu}^{C,\,
\beta}(x)=\frac{1}{2\pi} \int dk \hat{L}(k)e^{ikx}
\label{levydef}
 \ee
where  the characteristic function is given by
\be
\log \hat{L}(k)= -C|k|^{\mu} (1+i \beta ~{\rm sign}(k) ~\tan
(\pi \mu /2)).
\label{logchar}
\ee
The parameters $\mu$, $\beta$ and $C$ are related
to the asymptotic behavior of $L_{\mu}^{C,\beta}(x)$
\be
 \lim_{x \rightarrow \pm \infty}  L_{\mu}^{C ,\beta}(x)=
\gamma(\mu) \frac{C (1\pm \beta)}
{|x|^{\mu+1}}
\label{levy1}
\ee
with  the $\mu$-dependent parameter $\gamma(\mu)$ given by
\be
\gamma(\mu)=\Gamma(1+\mu)\sin(\frac{\pi\mu}{2})
\ee
Here $\mu$ is the stability index defined in the interval $(0,2]$,
$-1 \le\beta\le 1$ measures the asymmetry of the distribution
and the range $C > 0$ is the analogue of the variance, in a sense
that a typical value of $x$ is $C^{1/\mu}$. A standard choice
corresponds to $C=1$.

We shall consider here only the stability index in the range $(1,2)$,
although as will be shown later, results obtained in this range seem to be
valid also for $\mu=1$. We also assume that all random variables have zero
mean.

\subsection*{Determination of the eigenvalue density}

A method of calculating the eigenvalue density of the Wigner-L\'evy
matrices was invented by Bouchaud and Cizeau \cite{BC}. Let us 
in this section briefly recall the main steps of the method. 
It is convenient to introduce the resolvent, called also Green's 
function:
\be
g(z) =\frac{1}{N}\langle {\rm Tr} \; G(z) \rangle
\ee
where elements of the matrix $G(z)$ are
\be G_{ij}(z)=
(z-H)_{ij}^{-1}
\label{resolvent}
\ee
and the averaging is carried out using the measure (\ref{measure}).
The resolvent contains the same information as the eigenvalue density 
$\rho(\lambda)$. Indeed if one approaches the real axis one finds
that $\rho(\lambda) = -1/\pi \lim_{\epsilon \rightarrow 0^+}
{\rm Im} \; g(\lambda + i\epsilon)$. This is
how one usually calculates $\rho(\lambda)$ from $g(z)$. 
For the Wigner-L\'evy ensemble, where individual matrix
elements have large scale-free statistical fluctuations
a slightly different method turns out to be more practical --
a method which allows one to avoid problems in
taking the double limit (first $N\rightarrow \infty$ 
and $\epsilon \rightarrow 0^+$) in which the
fluctuations are suppressed in an uncontrollable way in
the presence of an imaginary part of $z$.
In the BC method $z$ is kept on the real axis and
fluctuations are not suppressed, so one can safely take
the large $N$ limit.

The method goes as follows \cite{BC}. 
One first generates a symmetric $N\times N$ random matrix $H$
using the measure (\ref{measure}) 
and then by inverting $H\!-\!z$ one
calculates the resolvent $G(z)$ (\ref{resolvent}).
Next one adds a new row (and a symmetric column) 
of independent numbers identically distributed as those in the old matrix $H$.
One obtains a new $(N\!+\!1)\times (N\!+\!1)$ matrix $H^{+1}$, 
where $+1$ emphasizes that it has one more row and 
one more column than $H$. 
It is convenient to number elements of the original
matrix $H_{ij}$ by indices running over the range
$i,j=1,\dots, N$, and assign the index $0$ to the new row and the new column 
so that now indices of $H^{+1}$ run over the range $0,\dots, N$.
If one inverts the matrix $H^{+1}\! - \!z$ one obtains a new 
$(N\!+\!1)\times (N\!+\!1)$ resolvent $G^{+1}(z)$.
One can show that it obeys a recursive relation~\cite{BC}
\be
z-\frac{1}{G_{00}^{+1}(z)}= H_{00} + \sum_{i,j}^N H_{0i}
H_{0j} G_{ij}(z) = \frac{h_{00}}{N^{1/\mu}} + \sum_{i,j}^N
\frac{h_{0i}h_{0j} G_{ij}(z)}{N^{2/\mu}}.
\label{diagoo}
\ee
which relates the element $G_{00}^{+1}(z)$ 
of the $(N\!+\!1)\times (N\!+\!1)$ resolvent to 
the elements of the old $N\times N$ resolvent $G(z)$.
One can also derive similar equations for off-diagonal elements
of $G^{+1}(z)$:
\be
\frac{G_{0j}^{+1}(z)}{G_{00}^{+1}(z)}=
\sum_i^N \frac{h_{0i}G_{ij}(z)}{N^{2/\mu}}
\label{off-diag}
\ee
The difference between the probability distribution of the 
elements of $G^{+1}(z)$ and of $G(z)$ dissapears 
in the limit $N\rightarrow \infty$. 
The diagonal elements
of the matrix $G^{+1}(z)$ are identically distributed as the diagonal 
elements of the matrix $G(z)$. The same holds
for off-diagonal ones. Moreover in this limit
all elements of the $G$ matrix
become independent of each other. In particular
all diagonal elements of $G(z)$ 
become independent identically distributed (iid)
random variables as $N\rightarrow \infty$.
One can use equations (\ref{diagoo}) and
(\ref{off-diag}) to derive self-consistency equations 
for the probability density function (pdf)
for diagonal and the pdf for off-diagonal
elements. We are here primarily interested in 
the distribution of the diagonal elements, as we shall
see below. The self-consistency
equation for the probability distribution of diagonal
elements follows from the equation (\ref{diagoo}) 
and is independent of the distribution of the off-diagonal elements.
This can be seen as follows.
Let us first define after Bouchaud and Cizeau~\cite{BC} 
a quantity:
\be
S_0(z) = z - \frac{1}{G^{+1}_{00}(z)}
\label{SG}
\ee
It is merely a convenient change of variables  suited
to the left hand side of equation (\ref{diagoo}).
It is clear that if one determines the pdf for $S=S_0(z)$,
one will also be able to determine
the pdf for $G=G^{+1}_{00}(z)$ since the two pdfs can
be obtained from each other by the 
change of variables (\ref{SG}):
\be
P_G(G) = \frac{1}{G^2} P_S\left(z-\frac{1}{G}\right)
\label{vchange}
\ee
where $P_G$ and $P_S$ are pdfs for $G^{+1}_{00}(z)$
and $S_0(z)$ respectively. Since the probability distribution
$P_G$ is identical for all diagonal elements of $G$,
it remains to determine 
the probability distribution $P_S$ for the quantity $S_0(z)$.

Let us sketch how to do that.
First observe that the first term in the equation (\ref{diagoo})
can be neglected at large $N$, so the equation assumes the
form:
\be
S_0(z)=  \sum_{i}^N
\frac{h^2_{0i} G_{ii}(z)}{N^{2/\mu}}
+\sum_{i\neq j}
\frac{h_{0i}h_{0j} G_{ij}(z)}{N^{2/\mu}}\,\,.
\label{SELF1}
\ee
Now observe that by construction $h_{0i}$ are independent
of $G_{ij}(z)$. We shall now show that for large $N$
the first term in (\ref{SELF1}) dominates over the
second one. We shall modify here the  
argument used in \cite{BC}, where it was assumed that the off-diagonal 
elements $G_{ij}N(z),~i\ne j$ are suppressed by a factor $1/N^{1/\mu}$
with respect to the diagonal
elements. Instead we note that by construction the quantities
$G_{ij}(z)$ are statistically independent of $h_{0i},~i=0,\dots,N$.
Since we shall only be interested by the
diagonal elements of the resolvent matrix, we may replace the contribution
of the off-diagonal elements $h_{0i}h_{0j}G_{ij}(z),~i\ne j$ by their
averaged values. As a result, the contribution of the off-diagonal terms
averages out. Note that this is also true in the Gaussian limit $\mu=2$
where following \cite{BC} we may replace all elements of (\ref{SELF1}) by
their respective averages. Taking this into account and omitting the
subleading contribution $H_{00}$, we get
\beq
S_0(z) = \sum_i^N \frac{h_{0i}^2 G_{ii}(z)}{N^{2/\mu}}.
\label{SELF2}
\eeq
Thus the problem was simplified to 
an equation where the left hand side ($S_0=z-1/G^{+1}_{00}$)
and the right hand side depend only of the diagonal elements of 
the $G$-matrix, which as we mentioned before, are 
identically distributed in the limit $N \rightarrow \infty$.
Using (\ref{SELF2}) one can derive a self-consistency equation for the 
probability density function (pdf) $P_G$ for diagonal elements
of the matrix $G$.

\subsection*{Generalized central limit theorem}

To proceed further we apply with Bouchaud and Cizeau~\cite{BC}
the {\it generalized central limit theorem} to derive
the universal behavior of the sum on the right hand side of
(\ref{SELF2}) in the limit $N\rightarrow \infty$:
\begin{itemize}
\item {\bf i.} If the $h_{0 i}$'s are sampled from
the L\'{e}vy distribution $L_{\mu}^{C,\beta}$, the squares $t_i=h_{0i}^2$
for large $t_i$ are distributed solely along the positive real axis,
with a heavy tail distribution:
\beq
\propto \gamma(\mu) \frac{C dt_i}{t_i^{1+\mu/2}}
\eeq
irrespective of $\beta$. The sum
\beq
\sum_i^N \frac{t_i}{N^{2/\mu}}
\eeq is distributed following
$L_{\mu/2}^{C',1}(t_i)$. The range parameters $C$ and
$C'$ are related by~\rf{logchar}
\beq
2 C' \gamma(\mu/2) = C \gamma(\mu)
\label{scaling}
\eeq
The factor 2 on the left hand side corresponds to the sum $(1+\beta)+(1-\beta)$
appearing as a contribution from positive and negative values of the original
distribution of $h_{0i}$.
This relation  is important when
comparing to the numerical results below where $C=1$ is used.
From now on (and to simplify the equations) we assume instead that $C'=1$.
\item {\bf ii.} By virtue of the central limit theorem 
the following sum 
\beq
\sum_i^N \frac{G_{ii}(z) t_i}{N^{2/\mu}}
\eeq
of iid heavy tailed numbers $t_i$ is 
for $G_{ii}(z)={\cal O}(N^0)$ and $N\rightarrow \infty$
L\'evy distributed with the pdf:
$L_{\mu/2}^{C(z),\beta(z)}$, which has the stability index $\mu/2$
and the effective range $C(z)$ and the asymmetry parameter $\beta(z)$
calculated from the equations:
\beq
C(z)=\frac{1}{N}\sum_i^N |G_{ii}(z)|^{\mu/2}
\label{ceff}
\eeq
and
\beq
\beta(z)=\frac{\frac{1}{N}\sum_i^N |G_{ii}(z)|^{\mu/2}{\rm sign}(G_{ii}(z))}
{\frac{1}{N}\sum_i^N |G_{ii}(z)|^{\mu/2}}\label{beta}\,\,.
\eeq
which follow from the composition rules for the tail amplitudes
of iid heavy tailed numbers $t_i$ defined above.
\end{itemize}

\subsection*{Integral Equations}

We saw in the previous section that 
the generalized central limit theorem implies for large
$N$ that the ``self-energy'' $S=S_0(z)$ is distributed according to the
L\'evy law $P_S(S) = L_{\mu/2}^{C(z),\beta(z)}(S)$ with the
stability index $\mu/2$ being a half of the stability index
of the L\'evy law
governing the distribution of individual elements
of the matrix $H$, and with the
effective range parameter $C(z)$ and the
asymmetry parameter $\beta(z)$ which can be calculated
from the equations (\ref{ceff}) and (\ref{beta}), respectively.
One should note that the effective parameters $C(z),\beta(z)$
of the distribution $P_S(S)$ are calculated
for $C=1$ and that they are independent of $\beta$ of
the probability distribution: $L_\mu^{C,\beta}(h_{ij})$ of 
the $H$-matrix elements.

The sums on the right hand side of the two equations 
(\ref{ceff}), (\ref{beta}) for $C(z)$ and $\beta(z)$ 
have a common form $\frac{1}{N} \sum_i f(G_{ii}(z))$.
Since in the limit $N\rightarrow \infty$, the diagonal
elements become iid, the sums on can be substituted
by integrals over the probability density for $G_{ii}$:
\be
\frac{1}{N}\sum_i^N \langle f(G_{ii}(z))\rangle = 
\int dG \; P_G(G) \; f(G)  = 
\int \frac{dG}{G^2} P_S\left (z-\frac{1}{G}\right) \; f(G)
\ee
where in the second step we used the equation (\ref{vchange}).
Since the distribution
$P_S(S) = L_{\mu/2}^{C(z),\beta(z)}(S)$ is known
up to the values of two effective parameters
$C(z)$ and $\beta(z)$
the equations (\ref{ceff}) and (\ref{beta}) can be written
as self-consistency relations for $\beta(z)$ and $C(z)$:
\bea
C(z) &=& \Pint\frac{dG}{G^2}
|G|^{\mu/2}L_{\mu/2}^{C(z),\beta(z)}
(z-1/G)\nonumber \\
\beta(z)&=&\frac{\pint\frac{dG}{G^2}
|G|^{\mu/2}{\rm sign}(G) L_{\mu/2}^{C(z),\beta(z)}(z-1/G)}
{\pint\frac{dG}{G^2} |G|^{\mu/2}
L_{\mu/2}^{C(z),\beta(z)}(z-1/G)}.\label{correct}
\eea
The symbol $-\hspace{-11pt}\int$ stands for principal value of the
integral.  Notice here the difference between our second equation
and that in~\cite{BC}. In addition to~\cite{BC} we also note
that the resolvent takes the form:
\beq
g(z) = \frac{1}{N} \sum_i G_{ii}(z) \ \longrightarrow \ 
g(z) = \Pint \frac{dG}{G^2}~G~ L_{\mu/2}^{C(z),\beta(z)} (z-1/G)
\label{G}
\eeq
The integrals \rf{correct} and \rf{G} can be rewritten using the
new integration variable $x=1/G$ as
\bea
C(z)&=&\int^{+\infty}_{-\infty} 
dx |x|^{-\mu/2}L_{\mu/2}^{C(z),\beta(z)}(z-x)\nonumber\\
\beta(z)&=&
\frac{\int^{+\infty}_{-\infty} 
dx ~{\rm sign}(x)|x|^{-\mu/2}L_{\mu/2}^{C(z),\beta(z)}(z-x)}
{\int^{+\infty}_{-\infty} dx |x|^{-\mu/2}L_{\mu/2}^{C(z),\beta(z)}(z-x)}
\label{correct1}
\eea
and
\beq
g(z)=\Pint \frac{dx}{x}L_{\mu/2}^{C(z),\beta(z)}(z-x)
\eeq
All the steps above require both $z$ and $G_{ii}(z)$
to be strictly real. The argument cannot be extended to the complex $z$ plane.
So all integrals above should be interpreted as principal value integrals, 
wherever it is necessary.
Since $\mu < 2$ all integrals in \rf{correct1} are convergent also
in the usual sense.

The equation for $g(z)$ can be rewritten as
\beq g(z)= \Pint\frac{dx}{z-x}L_{\mu/2}^{C(z),\beta(z)}(x).
\label{gdef1} \eeq 
Notice the nontrivial dependence on $z$ in the
parameters of the L\'{e}vy distribution. Above equation can be a
source of  confusion, since its structure  resembles another
representation of $g(z)$
\beq
g(z) =  \Pint \frac{d\lambda}{z-\lambda}\rho(\lambda)
\label{hilbert}
\eeq
which superficially looks as if one could identify in \rf{gdef1} $x$ and
$\lambda$ and $L_{\mu/2}^{C(z),\beta(z)}(x)$ with $\rho(\lambda)$. This is
not the case, and one should instead invert 
(\ref{gdef1}) using the inverse Hilbert transform:
\beq
\rho(\lambda) = \frac{1}{\pi^2} \Pint \frac{dz}{z-\lambda} g(z).
\label{disp}
\eeq
In other words, one first has to reconstruct numerically  the real
part of the resolvent, and only then  compute numerically the
spectral function  $\rho(\lambda)$, using the "dispersive relation"
(\ref{disp}). This is a difficult and rather subtle procedure.
In the next section we give some analytical insights to the
integral equations that would help solve them and extract the
spectral function.

\subsection*{Analytical properties useful for numerics}

One cannot do the integrals from previous section analytically.
As mentioned, one cannot even write down an explicit form
of the L\'evy distribution. It is a great numerical challenge
to solve the problem numerically even if all the expressions
are given. One realizes that already
when one tries to compute the Fourier integral (\ref{levydef})
of the characteristic function since one immediately
sees that the integrand in the form (\ref{levydef})
is a strongly oscillating function making the numerics
unstable. Fortunately using the power of the complex
analysis one can change this integral to a form which
is numerically stable. So in this section we present 
some analytic tricks which allow one to reduce the problem 
of computing the eigenvalue density 
as formulated in the previous section 
to a form which is well suited to 
the numerical computation.

To simplify \rf{correct1} we proceed in steps. First, we make
use of the L\'{e}vy distribution through its characteristic
\beq
L_{\mu/2}^{C,\beta}(x)=\frac{1}{2\pi}\int_{-\infty}^{\infty}dk
e^{ikx} e^{-C|k|^{\mu/2}(1+i\zeta ~{\rm sign}(k))}. \label{Levyf}
\eeq
with $\zeta=\beta~\tan(\pi\mu/4)$. By rescaling through

\bea k &=& C^{-2/\mu}k',\\
\nonumber x &=& C^{2/\mu}x', \\ \label{scale1} z &=& C^{2/\mu}z',
\nonumber
\eea
we can factor out the range
$L_{\mu/2}^{C,\beta}(x)=C^{-2/\mu}L_{\mu/2}^{1,\beta}(x')$.
Second, we make use of the following integrals,

\bea
\Pint \frac{dx}{z-x}e^{ikx}&=&-2ie^{ikz}{\rm sign}k,\nonumber \\
\int_0^{\infty}dx'~\frac{\cos(k'x')}{x'^{\mu/2}}&=&|k'|^{\mu/2-1}\Gamma(1-\mu/2)
\sin(\pi\mu/4),\\ \nonumber
\int_0^{\infty}dx'~\frac{\sin(k'x')}{x'^{\mu/2}}&=&|k'|^{\mu/2-1}{\rm
sign}(k') \Gamma(1-\mu/2)\cos(\pi\mu/4).
\eea
Last, we make use of the change of variables $p=k'^{\mu/2}$.
With this in mind, we obtain

\bea
C^2(z')&=&\frac{4}{\pi\mu}\Gamma\left(1-\frac{\mu}{2}\right)
\sin\left(\frac{\pi\mu}{4}\right) \int_0^{\infty} dp
\cos(p^{2/\mu}z'-\zeta(z')~ p)e^{-p},\\ \label{C}
\zeta(z')&=&
\frac{\int_0^{\infty} dp \sin(p^{2/\mu}z'-\zeta(z')~ p)e^{-p}}
{\int_0^{\infty} dp \cos(p^{2/\mu}z'-\zeta(z')~ p)e^{-p}}, \label{bet}
\eea
with $\zeta(z')=\tan(\pi\mu/4)~\beta(z')$.
For every $z'$ we can iteratively solve the equation for $\zeta(z')$,
then we determine $C(z')$ and use $z = C^{2/\mu}(z') z'$ to express
everything in terms of $z$. These transformations solve~\rf{correct1}.
Using the same method we rewrite the equation for $g(z)$ as

\beq
\bar{g}(z')=C(z')^{2/\mu}g(z) =\frac{2}{\mu}\int_0^{\infty}dp~p^{(2-\mu)/\mu}
\sin(p^{2/\mu}z' - p~\zeta(z'))~e^{-p}.
\eeq

These integral forms are useful to study the small-$z'$ limit. In this
case $\zeta(z')$ is an antisymmetric function of $z'$ and has an expansion
in powers of $z'$. Using
$\zeta(z')=k_1 z'+ {\cal O}(z'^3) $ we can recursively obtain
the coefficients of this expansion. The first term is
$k_1=\Gamma(1+2/\mu)/2$. Similarly $C(z')$ is a symmetric function in $z'$

\beq
C^2(z')=\frac{4}{\pi\mu}\Gamma\left(1-\frac{\mu}{2}\right)
\sin\left(\frac{\pi\mu}{4}\right)+{\cal O}(z'^2)
\eeq

For $|z'|$ large ($z' \to \pm \infty$) a different approach
is needed. In this case we  follow
Nolan~\cite{NOLAN} and treat the two integrals (numerator and
denominator) of \rf{bet} together, i.e. we  consider the integral

\beq \int_0^{\infty} dp e^{-h(p)}, \eeq where \beq h(p) =
(1-i\zeta)p + iz' p^{2/\mu}.
\eeq
Nolan's idea is to close the contour of integration in the
complex $p$ plane in the
following way: at $p \to \infty$ we add an arc and afterwards
continue until $p=0$ along the line where $\Im h(p)$=0. Using the
parameterization $p=re^{i\theta}$ we  get the parametric equation
for $r(\theta)$ along this line, valid for $z'>0$

\beq r(\theta)=\left(\frac{\sin(\theta_0-\theta)}{z' \cos(\theta_0)
\cos(\frac{2}{\mu}\theta)}\right)^{\mu/(2-\mu)}.
\eeq
The angle
$\theta$ for the curve we need ($\mu \ge 1$) is bounded between
$\theta_0 = {\rm arctan}(\zeta(z'))$, where $r=0$ and
$-\theta_1=-\pi\mu/4$, where $\cos(\frac{2}{\mu}\theta)$ is zero.
Let us now  introduce a new variable $\psi$ through
$\theta=\theta_0-\psi$ and $0 \le \psi \le \theta_0+\theta_1$. In
this range we have \bea
V(\psi)&=&\left(\frac{\sin(\psi)}{\cos(\theta_0)
\cos(\frac{2}{\mu}(\psi-\theta_0))} \right)^{\mu/(2-\mu)}, \\
\nonumber
\Re h(\psi)&=&\left(\frac{1}{z'}\right)^{\mu/(2-\mu)}V(\psi)
\frac{\cos(\frac{2-\mu}{\mu}\psi
-\frac{2}{\mu}\theta_0)}{\cos(\theta_0)
\cos(\frac{2}{\mu}(\psi-\theta_0))}. \eea
After some manipulations we obtain

\bea \int \sin(p^{2/\mu}z'-\zeta(z')~ p)e^{-p}
&=&\int_0^{\theta_0+\theta_1}d\psi
\left(\frac{1}{z'}\right)^{\mu/(2-\mu)}V(\psi)e^{-
\Re h(\psi)}\\ \nonumber
&\times&\left(\frac{2}{2-\mu}\frac{\cos(\frac{2-\mu}{\mu}(\psi-\theta_0))}
{\cos(\frac{2}{\mu}(\psi-\theta_0)}-\frac{\mu}{2-\mu}\frac{\sin\theta_0}
{\sin\psi}\right)\\ \nonumber \int \cos(p^{2/\mu}z'-\zeta(z')~ p)e^{-p}
&=&\int_0^{\theta_0+\theta_1}d\psi
\left(\frac{1}{z'}\right)^{\mu/(2-\mu)}V(\psi)e^{-
\Re h(\psi)}\\ \nonumber
&\times&\left(\frac{2}{2-\mu}\frac{\sin(\frac{2-\mu}{\mu}(\psi-\theta_0))}
{\cos(\frac{2}{\mu}(\psi-\theta_0))}+\frac{\mu}{2-\mu}\frac{\cos\theta_0}
{\sin\psi}\right). \label{integrals} \eea

The resulting integrals look complicated, however they contain
both the small-$z'$ and the large-$z'$ asymptotics. For $z' \to 0$
we have $\psi = z' p^{2/\mu-1}$, which reproduces the small-$z'$
expansion presented above. For $z' \to \infty$ we have $\psi =
\theta_0+\theta_1-u/z'^{\mu/2}$. Note that in this limit \beq
\cos(\frac{2}{\mu}(\psi-\theta_0))=\sin(\frac{2}{\mu}\frac{u}{z'^{\mu/2}})
\eeq and the $z'$ dependence in ${\rm Re}h(\psi)$ vanishes in
leading order.

The large-$z'$ asymptotics requires some work. For the
leading orders we have

\bea
\int \sin(p^{2/\mu}z'-\zeta(z')~ p)e^{-p}
&\approx&\Gamma(1+\mu/2)\sin\theta_1~(z')^{-\mu/2}\\ \nonumber \int \cos
(p^{2/\mu}z'-\zeta(z')~ p)e^{-p} &\approx&\Gamma(1+\mu/2)
\cos\theta_1~(z')^{-\mu/2}.
\eea
Thus for $z' \to \infty$ we have

\beq
\zeta(z')=\tan(\theta_1)+{\cal O}(1/z'^{\mu/2}) \eeq and \bea
C(z')&=& z'^{-\mu/4}(1+{\cal O}(1/z'^{\mu/2}))\\ \nonumber z
&=&\sqrt{z'}(1+{\cal O}(1/z'^{\mu/2})).
\eea
All the formulas above apply to the case $z'>0$. One can also derive similar
formulas for $z'<0$, it is however more practical to use the symmetry
properties of the functions $C(z')$ and $\zeta(z')$.

As a final check let us compute $\bar{g}(z')$. A rerun of the
above transformations on the integrals give

\bea
\bar{g}(z')&=&\frac{2}{2-\mu}\int_0^{\theta_0+\theta_1}d\psi
\left(\frac{1}{z'}\right)^{2/(2-\mu)}V(\psi)^{2/\mu}e^{-{\rm
Re}h(\psi)}
\\ \nonumber
&\times&\left(\frac{2}{\mu}\frac{1}{\cos(\frac{2}{\mu}(\psi-\theta_0))}
+\frac{\sin(\frac{2-\mu}{\mu}\psi-\frac{2}{\mu}\theta_0)}
{\sin\psi}\right). \eea
Notice that both asymptotics follow from this representation.
For $z' \to \infty$ we have $\bar{g}(z')=1/z' + \cdots$.
which implies $g(z) = 1/z +\cdots$. This can be viewed as a check
of the correct normalization of the eigenvalue density distribution.

\begin{figure}
\psfrag{L}{\bf{\Large $\lambda$}}
\psfrag{R}{\bf{\Large $\rho(\lambda)$}}
\centerline{\scalebox{0.6}{\rotatebox{0}{\includegraphics{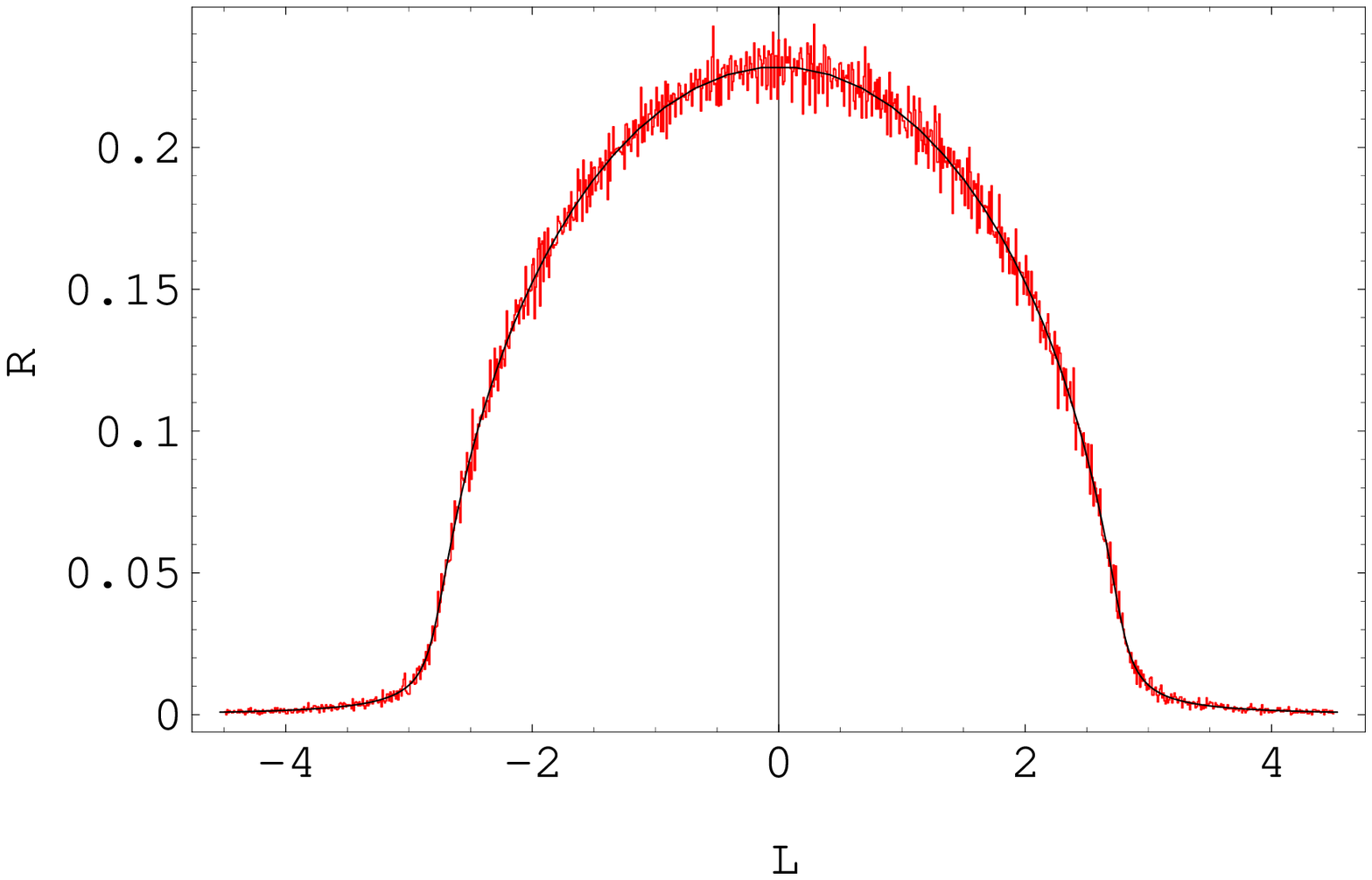}}}}
\caption[phased]{{\small Theoretical (black)
and numerical (red) eigenvalue distributions for $\mu=1.95$}}
\label{fig1}
\end{figure}
\begin{figure}
\psfrag{L}{\bf{\Large $\lambda$}}
\psfrag{R}{\bf{\Large $\rho(\lambda)$}}
\centerline{\scalebox{0.6}{\rotatebox{0}{\includegraphics{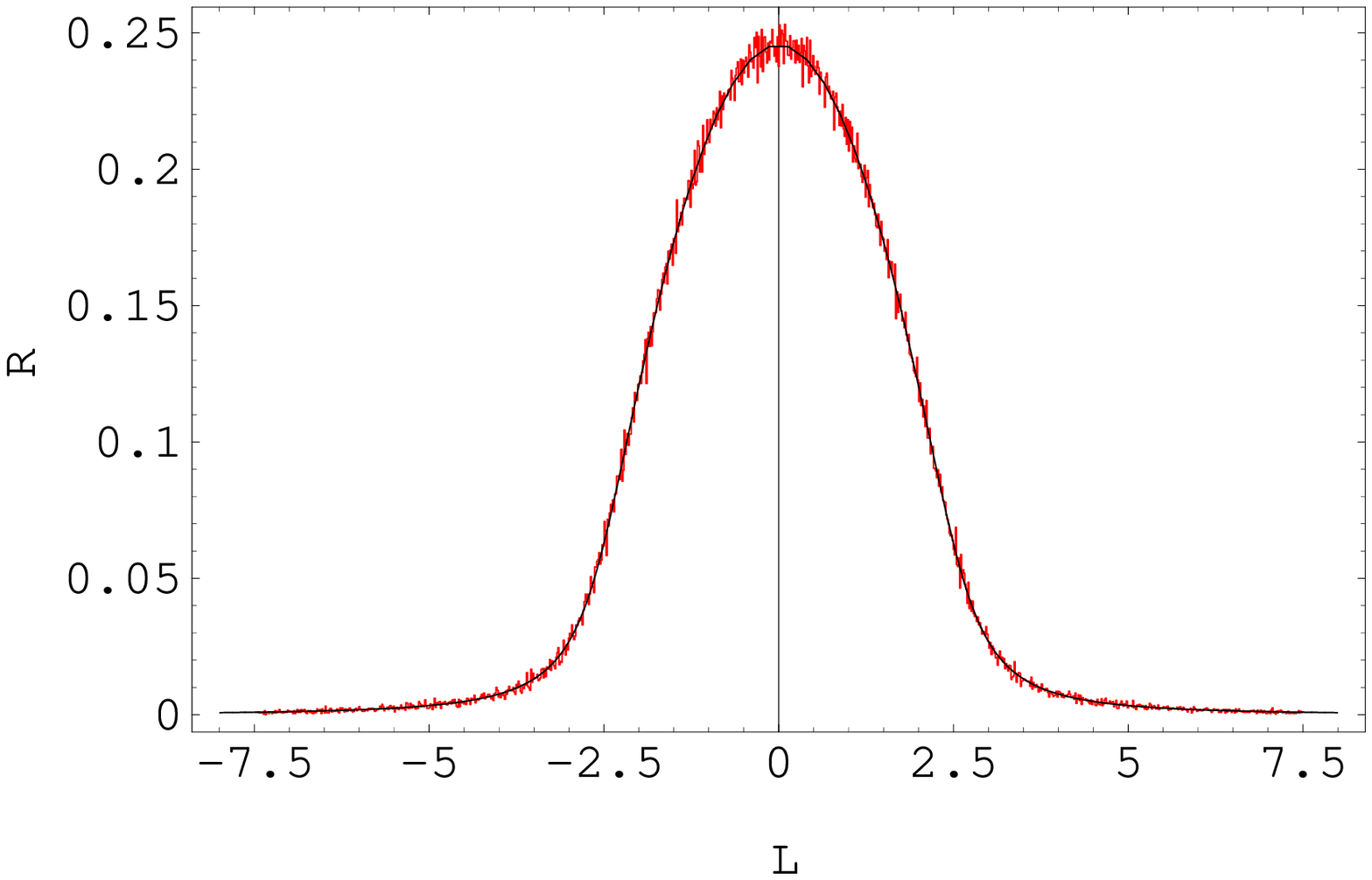}}}}
\caption[phased]{{\small Theoretical (black)
and numerical (red)  eigenvalue distribution for $\mu=1.75$}} \label{fig2}
\end{figure}
\begin{figure}
\psfrag{L}{\bf{\Large $\lambda$}}
\psfrag{R}{\bf{\Large $\rho(\lambda)$}}
\centerline{\scalebox{0.6}{\rotatebox{0}{\includegraphics{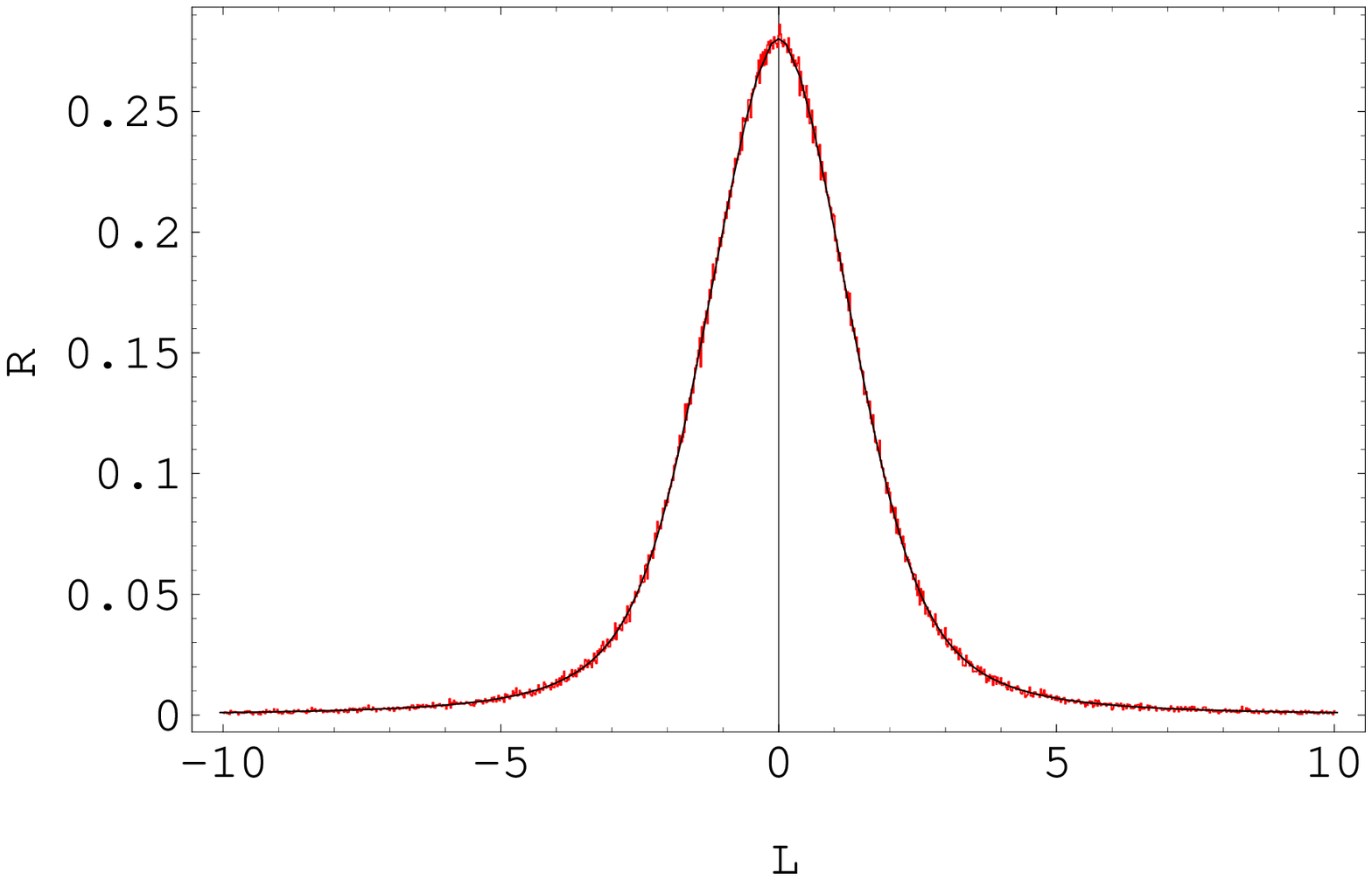}}}}
\caption[phased]{{\small Theoretical (black)
and numerical (red) eigenvalue distribution for $\mu=1.50$}} \label{fig3}
\end{figure}
\begin{figure}
\psfrag{L}{\bf{\Large $\lambda$}}
\psfrag{R}{\bf{\Large $\rho(\lambda)$}}
\centerline{\scalebox{0.6}{\rotatebox{0}{\includegraphics{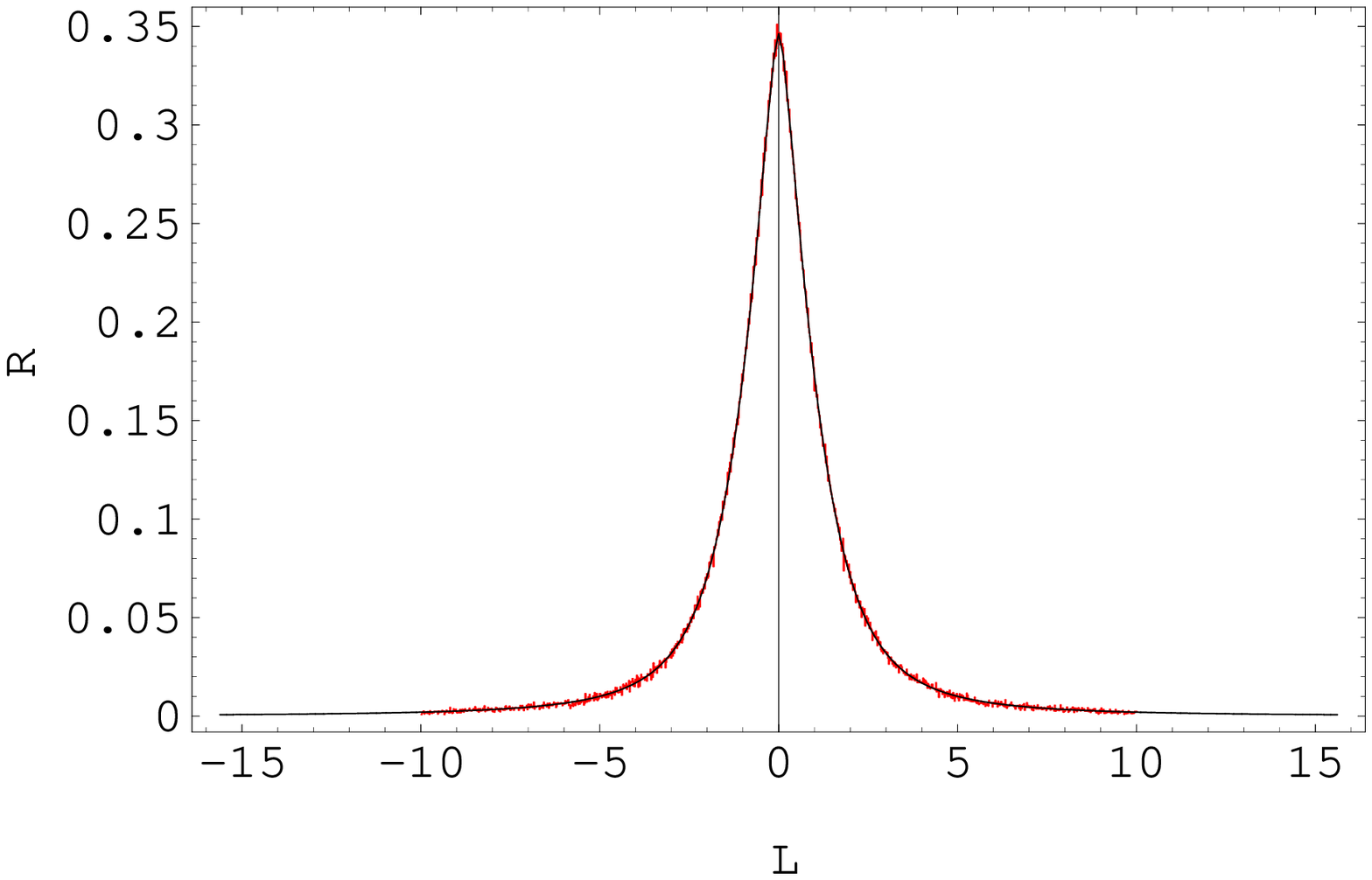}}}}
\caption[phased]{{\small Theoretical (black)
and numerical (red) eigenvalue distribution for $\mu=1.25$}} \label{fig4}
\end{figure}
\begin{figure}
\psfrag{L}{\bf{\Large $\lambda$}}
\psfrag{R}{\bf{\Large $\rho(\lambda)$}}
\centerline{\scalebox{0.6}{\rotatebox{0}{\includegraphics{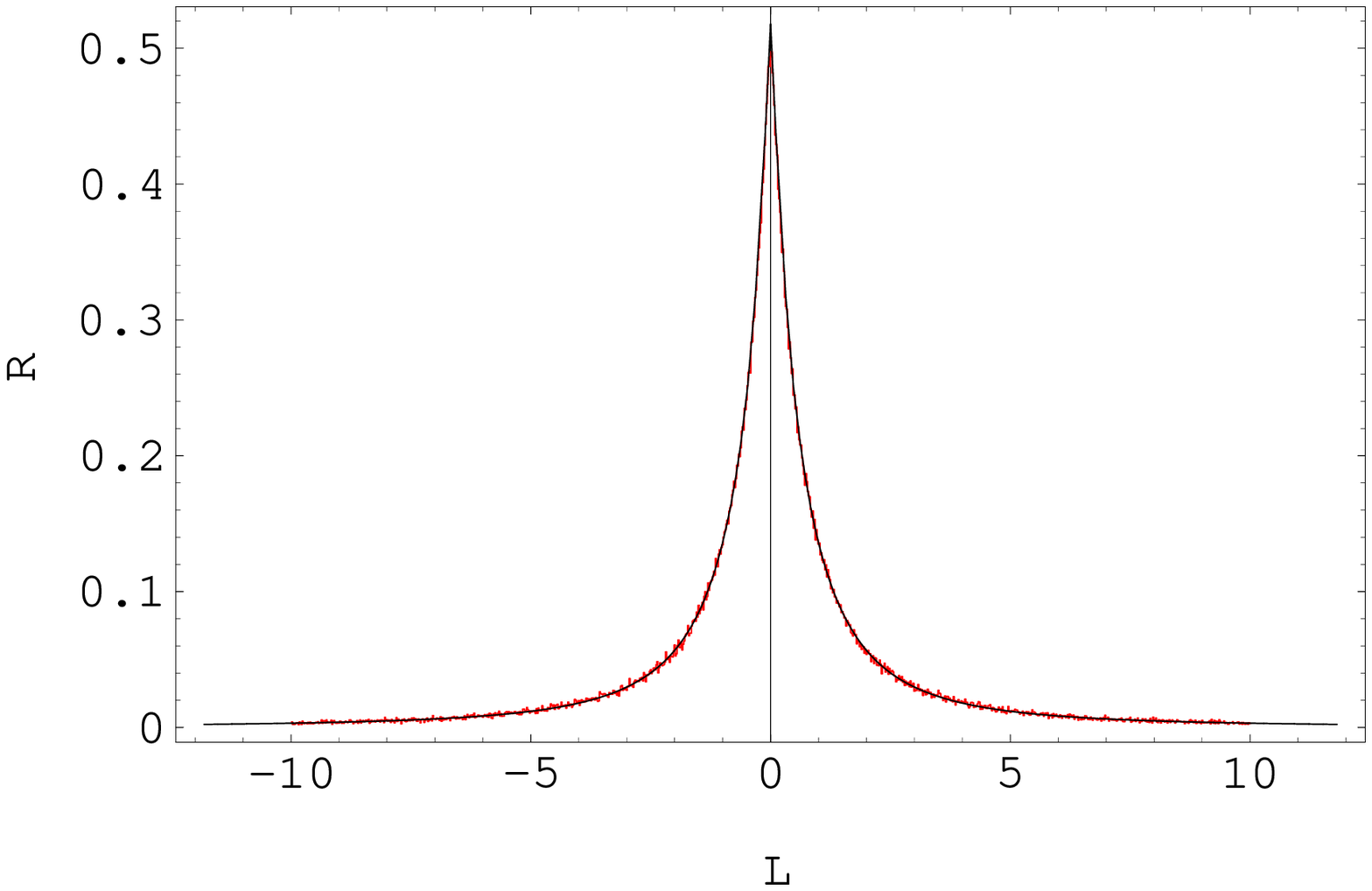}}}}
\caption[phased]{{\small Theoretical (black)
and numerical (red) eigenvalue distribution for $\mu=1.00$}} \label{fig5}
\end{figure}

\subsection*{Numerical Comparison}

In this section we show perfect agreement between the theoretical
analysis for the eigenvalue distribution~\rf{disp}
and the numerically generated eigenvalue distribution. The latter
is obtained by diagonalizing $N\times N$ random L\'{e}vy matrices
sampled using the measure~\rf{measure}. The former was generated
by calculating numerically $g(z)$ as detailed above.
We performed numerically the inverse Cauchy transform~\rf{disp}.
It is important to note that the integral transforms entering into
the definition of the eigenvalue density $\rho(\lambda)$ in
\rf{disp} converge slowly for $z\to \infty$. For that, we have used
the asymptotic expansion of $g(z)$ to perform the large-$z$ part of
the integrals.

As noted above, all the above analytical results were obtained
using a specific choice of the scale factor~\rf{scaling}. The
comparison with the numerically generated eigenvalue distribution
generated using $L_{\mu}^{1,0}$ distribution requires rescaling
through $\lambda \to \phi\lambda$ and
$\rho(\lambda)\to\rho(\lambda)/\phi$ with
\beq \phi=
\left(\frac{\Gamma(1+\mu)
\cos(\frac{\pi\,\mu}{4})}{\Gamma(1+\frac{\mu}{2})}\right)^{1/\mu}
\label{phi}
\eeq
Below we  show a sequence of results for
$\mu=1.95,~1.75,~1.50,~1.25$ and $1.00$ with this rescaling.
The comparison is for high statistics $400\times 400$ samples (red).
We have checked that the convergence is good already for $100\times 100$
samples, with no significant difference between $N=100, 200, 400$.
The numerical results are also not sensitive to the choice $\beta\neq 0$.
The agreement between the results following from the integral equations
and the numerically generated spectra is perfect.
This is true even for $\mu=1$, where in principle the arguments used in
the derivation may not be valid.

\subsection*{Numerical observation}

\begin{figure}
\psfrag{M}{\bf{\Large $\mu$}}
\psfrag{RH}{\bf{\Large $\rho(0)$}}
\centerline{\scalebox{0.6}{\rotatebox{0}{\includegraphics{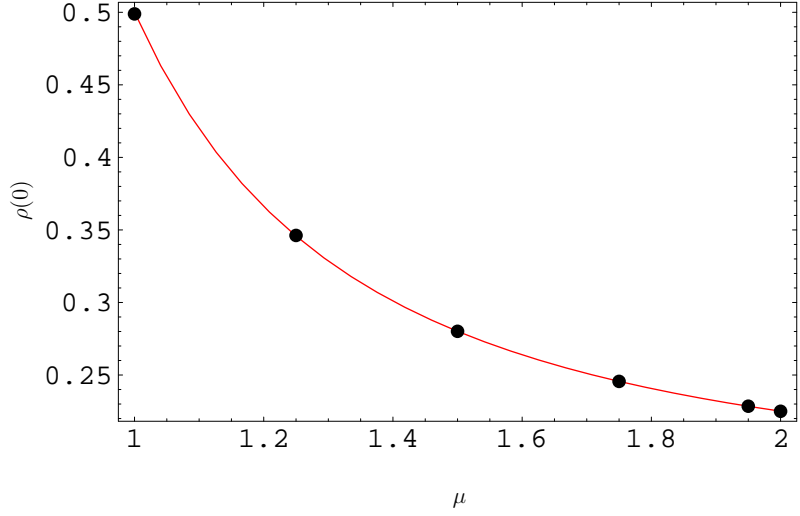}}}}
\caption[phased]{{\small The line represents the function
$\widehat{\rho}_\mu(0)$ (\ref{r0bc}) while the circles the values of  
$\rho_\mu(0)$ computed numerically for $\mu=1.0,1.25,1.5,1.75,1.95,2.0$ 
from the equation (\ref{disp}).}} 
\label{fig6}
\end{figure}

In the mean-field approximation \cite{BC} one can assume that
there are no correlations between large eigenvalues of the 
Wigner-L\'evy random matrix. In this case the eigenvalue 
density takes the form:
\beq 
\widehat{\rho}_\mu(\lambda) = L_{\mu/2}^{C(\lambda),\beta(\lambda)}(\lambda).
\label{gdef1hat} 
\eeq 
It is natural to ask how good this mean-field approximation is.
This can be done by comparing the mean-field eigenvalue 
distribution $\widehat{\rho}(\lambda)$ (\ref{gdef1hat})
to the eigenvalue distribution $\rho(\lambda)$ calculated 
by the inverse Hilbert transform (\ref{disp}) of 
the resolvent $g(z)$ (\ref{gdef1}) as we did
in previous section. We made this comparison numerically. 
The result of this numerical experiment was that
within the numerical accuracy which we achieved 
the two curves representing $\widehat{\rho}(\lambda)$
and $\rho(\lambda)$ lied on top of each other 
in the whole studied range of $\lambda$. Since our numerical
codes are written in {\em Mathematica} we could push the 
numerical accuracy very far, being only limited by the
execution time of the code. We have not seen any sign 
of deviation between the shapes of the two curves. 
This provides us with a strong numerical evidence that
the mean-field argument \cite{BC} gives 
an exact result but so far we have not managed to prove it.
The value of the eigenvalue density $\widehat{\rho}_\mu(0)$ 
for $\lambda=0$ can be calculated analytically for the mean-field
density (\ref{gdef1hat}). Rescaling the density 
$\widehat{\rho}_\mu(\lambda) \rightarrow \widehat{\rho}_\mu(\phi \lambda)/\phi$ 
by the factor $\phi$ (\ref{phi}) we eventually obtain
\begin{equation}
\widehat{\rho}_\mu(0) = 
\frac{\Gamma(1+2/\mu)}{\pi}\left(\frac{\Gamma(1+\mu/2)^2}{\Gamma(1+\mu)}\right)^
{1/\mu}
\label{r0bc}
\end{equation}
We draw this function in Fig.\ref{fig6}. In the same figure 
we also show points representing numerically evaluated values of 
the corresponding density $\rho_\mu(0)$ (\ref{disp}) at some values
of $\mu$. Within the numerical accuracy $\rho_\mu(0)$ and 
$\widehat{\rho}_\mu(0)$ assume the same values.

The physical meaning of the mean-field argument
is that indeed one can think of the large eigenvalues as 
independent of each other. A similar observation has
been made recently \cite{SF}. The mathematical meaning 
of the mean-field argument 
is more complex as we shall discuss below.

Let us for brevity denote the function 
$L_{\mu/2}^{C(z),\beta(z)}(x)$, which is a function 
of two real arguments, by 
$f(z,x) \equiv L_{\mu/2}^{C(z),\beta(z)}(x)$.
The equation (\ref{gdef1}) can be now written as:
\beq 
g(z)= \Pint d x \frac{f(z,x)}{z-x}
\label{gdef2} 
\eeq 
and (\ref{disp}) as
\beq
\rho(\lambda) = \frac{1}{\pi^2} \Pint dz \frac{g(z)}{z-\lambda} .
\eeq 
When we insert (\ref{gdef2}) into the last equation we have:
\beq
\rho(\lambda) = 
\frac{1}{\pi^2} \Pint dz \Pint dx \frac{f(z,x)}{(z-\lambda)(z-x)}.
\eeq
A question is when this exact expression for $\rho(\lambda)$ is
equal to the mean-field solution: 
$\widehat{\rho}(\lambda) = f(\lambda,\lambda)$.
Recall the Poincar\'e-Bertrand theorem.
It tells us that the following equation holds
\beq
f(\lambda,\lambda) = 
\frac{1}{\pi^2} \Pint dz \Pint dx \frac{f(z,x)}{(z-\lambda)(z-x)} -
\frac{1}{\pi^2} \Pint dx \Pint dz \frac{f(z,x)}{(z-\lambda)(z-x)}.
\label{pb}
\eeq 
We see that the density $\rho(\lambda)$ is given by the mean-field
result: $\rho(\lambda)=\widehat{\rho}(\lambda) \equiv f(\lambda,\lambda)$
if the second term on the right-hand side of (\ref{pb}) vanishes.
Unfortunately we have not managed to show that this is really
the case for $f(z,x) = L_{\mu/2}^{C(z),\beta(z)}(x)$.
One can however trivially observe that it would be the case if
$f(z,x)$ had the following form 
$f(z,x)=f(x,x)=L_{\mu/2}^{C(x),\beta(x)}(x)$,
and probably also if
$f(z,x)$ were a slowly varying function of $z$ for $z$ 
close to $x$, in which case the integral (\ref{gdef2}) would
pick up only the contribution from $f(x,x)$ leading to
(\ref{gdef1hat}).

\section{Free Random L\'{e}vy Matrices}

\subsection*{Rotationally invariant measure}

Clearly Wigner L\'evy matrices are not rotationally invariant.
In this section we shall discuss orthogonally (or unitary) invariant
ensembles of L\'evy matrices. It can be shown that maximally
random measures for such matrices have the form \cite{FREELEVY,BAL}:
\be
d\mu_{FR}(H)=\prod_{i\le j}\,dH_{ij}\,e^{-\,N{\rm Tr}V(H)}
\label{frmeasure}
\ee
We shall be interested only in potentials with have tails
which lead to eigenvalue distributions (spectral densities)
with heavy tails $\rho(\lambda) \sim \lambda^{-1-\mu}$ 
belonging to the L\'{e}vy domain of attraction.
A generic form of $V(\lambda)$ at asymptotic eigenvalues $\lambda$
is in this case
\be
V(\lambda)={\rm ln}\lambda^2+{\cal O}(1/\lambda^\mu)
\ee
In general the potential does not have to be an analytic function.
We shall be interested here only in stable ensembles in
the sense that the spectral measure~\rf{frmeasure} 
for the convolution of two independent and identical ensembles 
has the same form as the measure of the individual
ensembles. In other words, the spectral measure for
a matrix constructed as a sum of two independent matrices
taken from the ensemble has exactly the same spectral measure
(eigenvalue density) modulo linear transformations.

It turns out that one can classify all the stable spectral
measures thanks to the relation of the problem to free probability
calculus. The matrix ensemble (\ref{frmeasure}) is in the large
$N$ limit a realization of free random variables
\cite{FREELEVY}, so one can use theorems developed in free random
probability~\cite{BERVOIC}. In particular we can use the fact that
in free probability theory stable laws are classified. They actually
parallel stable laws (\ref{logchar})
of classical probability theory.
In free probability the analogue of the {\em logarithm} of the
characteristic function (\ref{logchar}) is the
R-transform, introduced  by Voiculescu~\cite{VOICULESCU}.
The R-transform linearizes the matrix convolution, generating 
spectral cumulants, which are additive under convolution.
\def\be{\begin{eqnarray}}
\def\ee{\end{eqnarray}}
\def\bea{\be}
\def\eea{\ee}
\def\la{\langle}
\def\ra{\rangle}
\def\half{{\textstyle \frac{1}{2}}}
\def\calO{{\cal O}}
\def\calB{{\cal B}}
\newcommand{\e}{{\mbox{e}}}
\def\del{\partial}
\def\vr{{\vec r}}
\def\vk{{\vec k}}
\def\vq{{\vec q}}
\def\vp{{\vec p}}
\def\vP{{\vec P}}
\def\vt{{\vec \tau}}
\def\vs{{\vec \sigma}}
\def\vJ{{\vec J}}
\def\vB{{\vec B}}
\def\hatr{{\hat r}}
\def\hatk{{\hat k}}
\def\roughly#1{\mathrel{\raise.3ex\hbox{$#1$\kern-.75em%
\lower1ex\hbox{$\sim$}}}}
\def\lsim{\roughly<}
\def\gsim{\roughly>}
\def\fm{{\mbox{fm}}}
\def\vx{{\vec x}}
\def\EM{{\rm EM}}
\def\bfpi{{\mbox{\boldmath $\pi$}}}
\def\bfrho{{\mbox{\boldmath $\rho$}}}
\def\bfalpha{{\mbox{\boldmath $\alpha$}}}
\def\bfSigma{{\mbox{\boldmath $\Sigma$}}}
\def\bfPi{{\mbox{\boldmath $\Pi$}}}
\def\bfGamma{{\mbox{\boldmath $\Gamma$}}}
\def\Tr{{\rm Tr}\,}
\def\barp{{\bar p}}
\def\zz{{z \bar z}}
\def\mus{{\cal M}_s}
\def\abs#1{{\left| #1 \right|}}
\def\ve{{\vec \epsilon}}
\def\nlo#1{{\mbox{N$^{#1}$LO}}}
\def\MS{{\mbox{M1V}}}
\def\mut{{\mbox{M1S}}}
\def\Qt{{\mbox{E2S}}}
\def\rM{{\cal R}_{\rm M1}}\def\rE{{\cal R}_{\rm E2}}
\def\J#1#2#3#4{ {#1} {\bf #2} (#4) {#3}. }
\def\PRL{Phys. Rev. Lett.}
\def\PL{Phys. Lett.}
\def\PLB{Phys. Lett. B}
\def\NP{Nucl. Phys.}
\def\NPA{Nucl. Phys. A}
\def\NPB{Nucl. Phys. B}
\def\PR{Phys. Rev.}
\def\PRC{Phys. Rev. C}
\def\MM{{\cal M}}
\newcommand{\eq}{\begin{equation}}
\newcommand{\eqx}{\end{equation}}
\newcommand{\eqn}{\begin{eqnarray}}
\newcommand{\eqnx}{\end{eqnarray}}
\newcommand{\f}[2]{\frac{#1}{#2}}
\newcommand{\lm}{\lambda}
\newcommand{\Lm}{\Lambda}
\newcommand{\lra}{\longrightarrow}
\newcommand{\tr}{\mbox{\rm tr}}
\newcommand{\al}{\alpha}
\newcommand{\bt}{\beta}
\newcommand{\eps}{\varepsilon}
\newcommand{\dl}{\delta}

\subsection*{Stable laws in free probability}
The remarkable achievement by Bercovici and Voi\-cu\-les\-cu~\cite{BERVOIC}
is an explicit derivation of all R-transforms defined by the equation
$R(G(z))=z-1/G(z)$ where $G(z)$ is the resolvent  for all free
stable distributions. 
We just note that $R(G(z))$ is a sort of self-energy for rotationally
symmetric FRL ensembles which is the analogue of (\ref{SELF2}) which
we previously defined for Wigner-L\'evy ensembles. 
It is self-averaging and additive.

For stable laws $R(z)$ is known. It 
can has either the trivial form $R(z)=a$ or
\be
R(z) = b z^{\mu-1}
\ee
where $0<\mu<2$, $b$ is a parameter which can be related to the
stability index $\mu$, the asymmetry parameter $\beta$, and the range $C$ 
known from the corresponding stable laws (\ref{logchar}) of
classical probability \cite{BERVOIC,PATA}
\indent{\be
b = \left\{ \begin{array}{cl}
C\ e^{i(\frac{\mu}2-\!1)(1\!+\!\beta)\pi}
     {\rm ~~for~~} 1 <\mu<2 \nonumber \\
 C \ e^{i[\pi+\frac{\mu}2(1\!+\!\beta)\pi]}
      {\rm ~~for~~} 0 <\mu <1 \nonumber
    \end{array} \right. \, .
\ee}
In the marginal case: $\mu=1$, $R(z)$ reads:
\be
R(z) = - i C(1+\beta) -
\frac{2\beta C}{\pi} \ln C z
\ee
The branch cut structure of $R(z)$ is chosen in such a way that the
upper complex half plane is mapped to itself.
Recalling that $R=z-1/G$ in the large $N$ limit,
one finds that for the trivial case $R(z)=0$,
the resolvent: $G(z)=z^{-1}$
and the spectral distribution
is a Dirac delta, $\rho(\lambda)=\delta(\lambda)$.
Otherwise, on the upper half-plane, the resolvent fulfills
an algebraic equation
\be
bG^{\mu}(z)-zG(z)+1=0\,\,,
\label{Levygreen}
\ee
or in the marginal case ($\mu=1$):
\be
\bigg(\!z\!+i C(1\!+\!\beta)\!\bigg) G(z) +\frac{2\beta C}{\pi}
G(z)\ln C G(z) - 1\! = \!0.
\ee
On the lower half-plane $G(\bar{z})=\bar{G}(z)$ \cite{BERVOIC}.

The equation for the resolvent (\ref{Levygreen})
has explicit solutions only for the following values:
$\mu=1/4,1/3,1/2, 2/3,3/4,4/3,3/2$ and $2$.
In all other cases the equation is transcendental and one
has to apply numerical procedures to unravel the spectral distribution.
Again the form of the potential generating
stable free L\'{e}vy ensembles is highly non-trivial and is
only known in few cases~\cite{FREELEVY}. We refer to~\cite{FREELEVY}
for further references and discussions.

\subsection*{Comparison of free L\'evy and Wigner-L\'evy spectra}

\begin{figure}
\psfrag{L}{\bf{\Large $\lambda$}}
\psfrag{R}{\bf{\Large $\rho(\lambda)$}}
\centerline{\scalebox{0.6}{\rotatebox{0}{\includegraphics{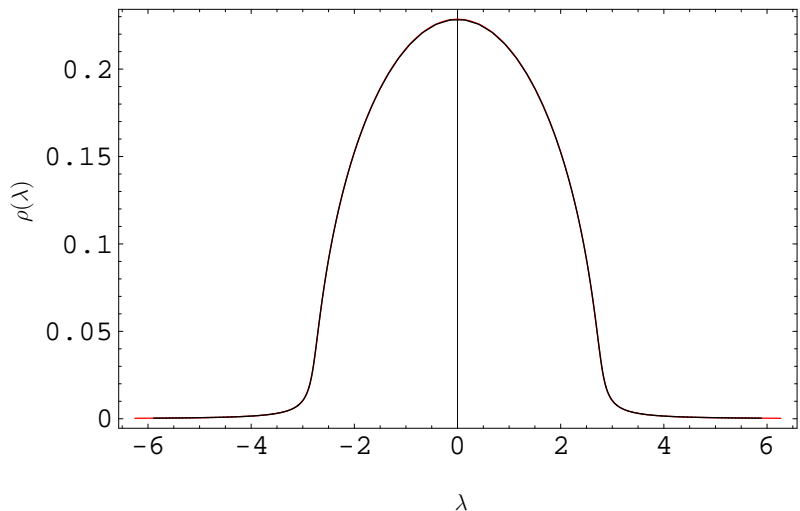}}}}
\caption[phased]{{\small WL (black) versus FRL (red)
for $\mu=1.95$}} \label{fig7}
\end{figure}
\begin{figure}
\psfrag{L}{\bf{\Large $\lambda$}}
\psfrag{R}{\bf{\Large $\rho(\lambda)$}}
\centerline{\scalebox{0.6}{\rotatebox{0}{\includegraphics{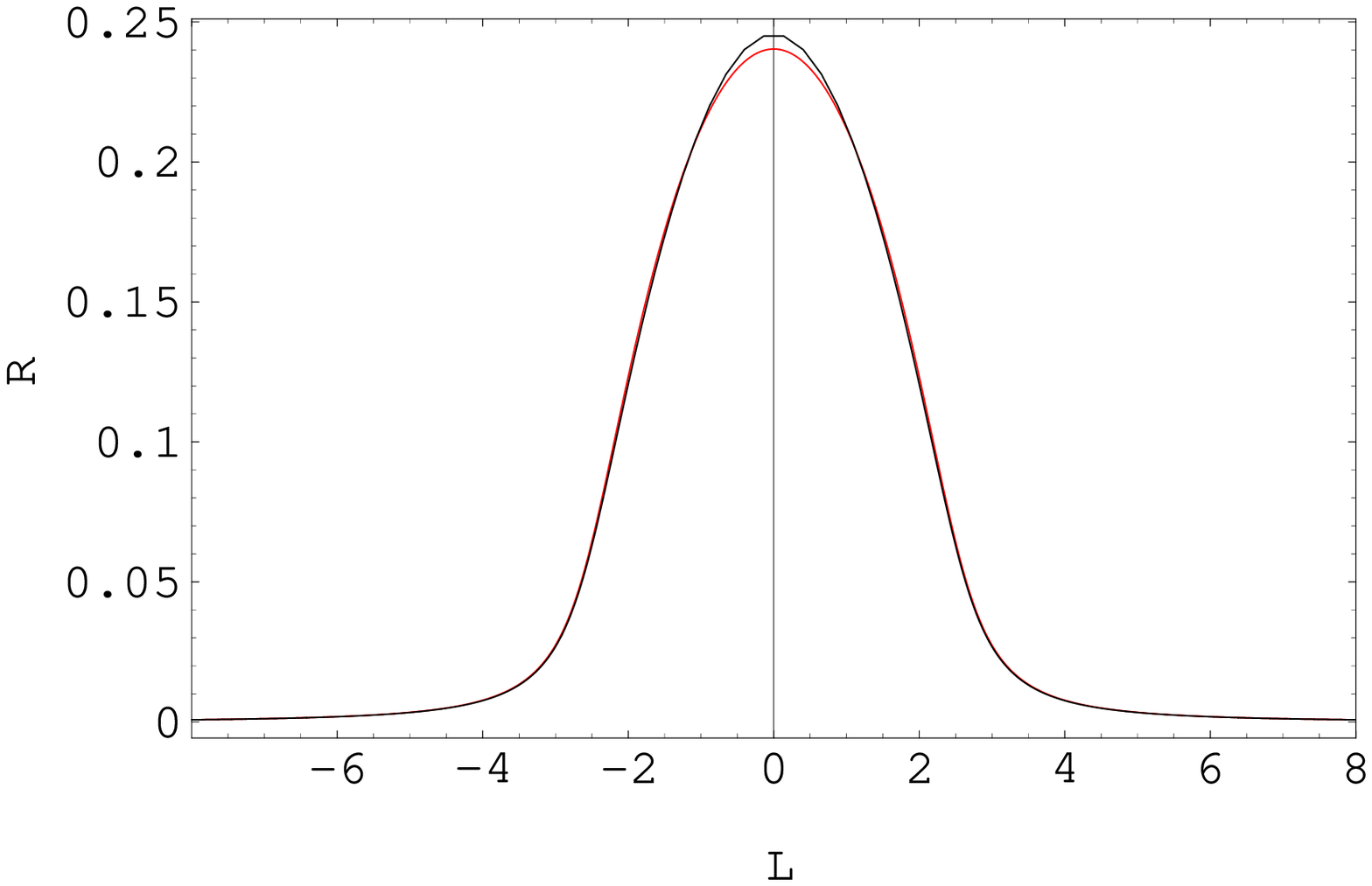}}}}
\caption[phased]{{\small WL (black) versus FRL (red)
for $\mu=1.75$}} \label{fig8}
\end{figure}
\begin{figure}
\psfrag{L}{\bf{\Large $\lambda$}}
\psfrag{R}{\bf{\Large $\rho(\lambda)$}}
\centerline{\scalebox{0.6}{\rotatebox{0}{\includegraphics{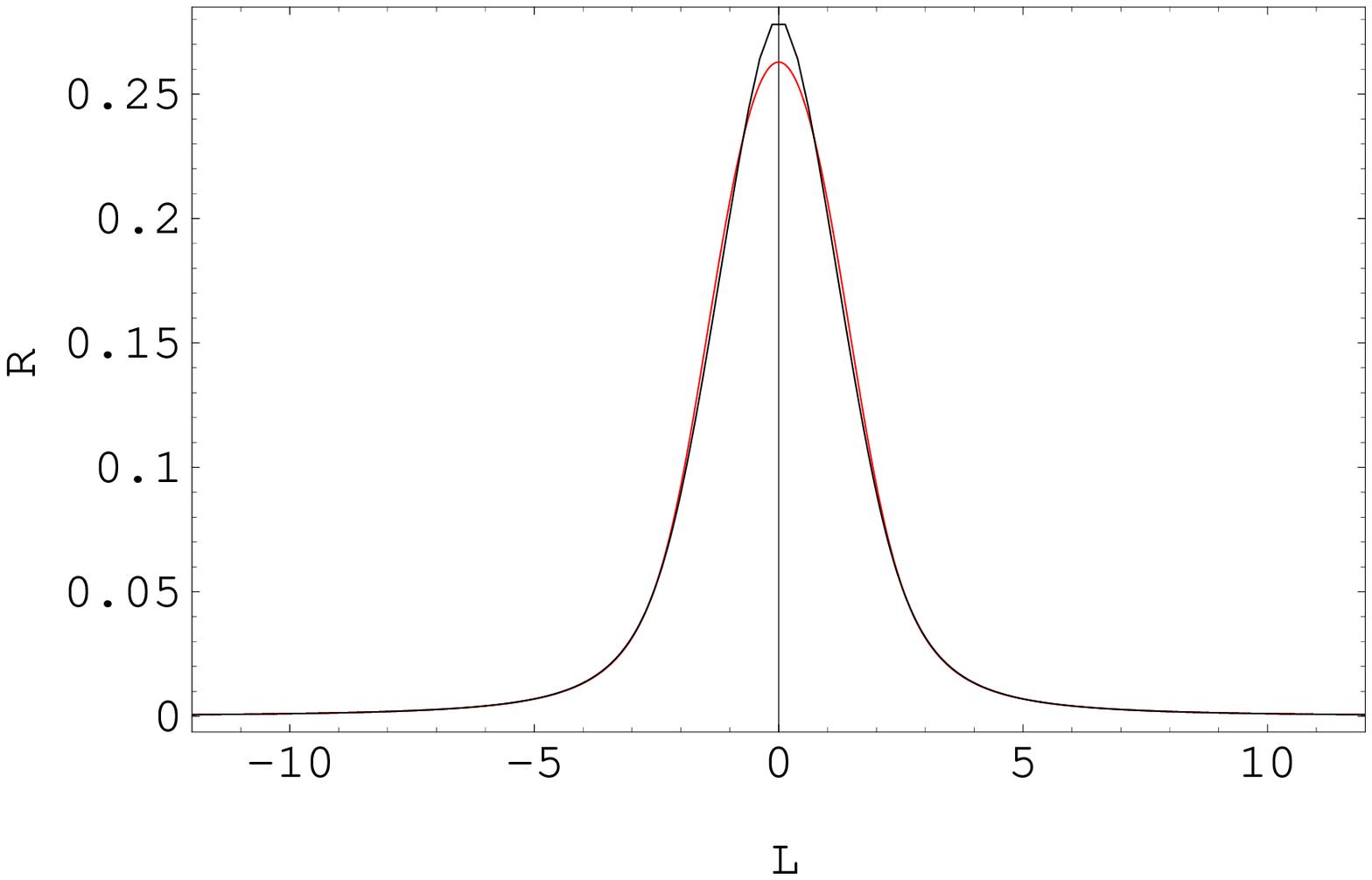}}}}
\caption[phased]{{\small WL (black) versus FRL (red)
for $\mu=1.50$}} \label{fig9}
\end{figure}
\begin{figure}
\psfrag{L}{\bf{\Large $\lambda$}}
\psfrag{R}{\bf{\Large $\rho(\lambda)$}}
\centerline{\scalebox{0.6}{\rotatebox{0}{\includegraphics{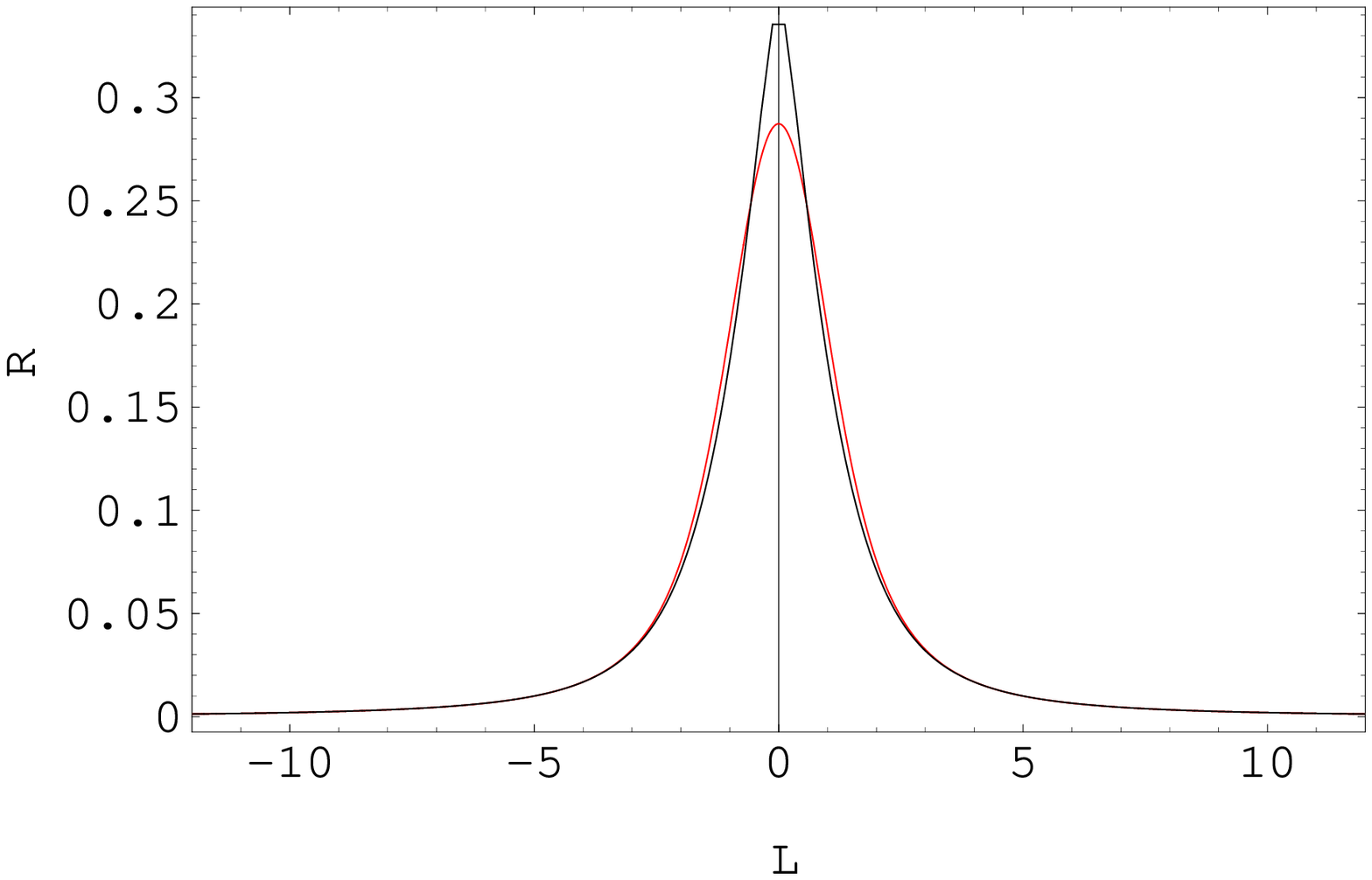}}}}
\caption[phased]{{\small WL (black) versus FRL (red)
for $\mu=1.25$}} \label{fig10}
\end{figure}
\begin{figure}
\psfrag{L}{\bf{\Large $\lambda$}}
\psfrag{R}{\bf{\Large $\rho(\lambda)$}}
\centerline{\scalebox{0.6}{\rotatebox{0}{\includegraphics{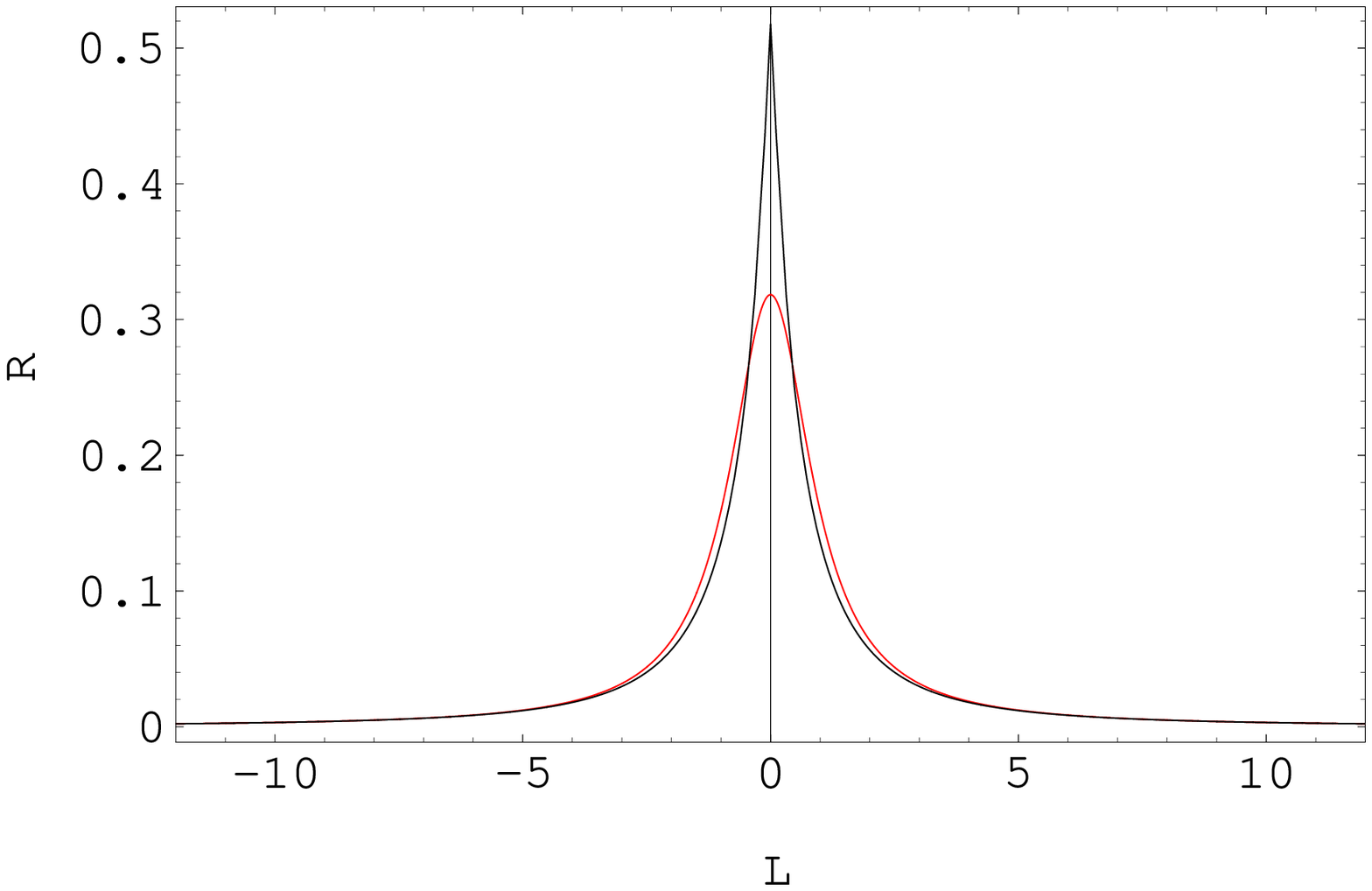}}}}
\caption[phased]{{\small WL (black) versus FRL (red)
for $\mu=1.00$}} \label{fig11}
\end{figure}
We present in Figures 7-11 several comparisons between
the free random L\'{e}vy spectra (FRL) following from the
solution to the transcendental equation (red) and the
random L\'{e}vy spectra (BC) obtained by solving the coupled
integral equations (black), for zero asymmetry ($\beta=0$)
and different tail indices $\mu$. The FRL spectra are normalized
to agree with the BC spectra in the tails of the distributions. 
We recall that the FRL spectra asymptote
$\rho(\lambda)\approx {\rm sin}(\pi \mu/2)/\pi$. 
The comparison in bulk shows that the spectra are similar, in particular 
close to the Gaussian limit $\mu=2$, where both approaches become equivalent.
For smaller $\mu$ there are differences.

WL and FRL matrices represent two types of random matrices 
spectrally stable under matrix
addition. For the WL matrices it follows from the measure, since each
matrix elements is generated from a stable L\'{e}vy distribution and therefore
the sum of $\cal N$ WL matrices, scaled by $1/{\cal N}^{1/\mu}$ is equivalent 
to the original WL ensemble. The important point is that the WL measure is
not symmetric while the FRL one is. 

\section{Spectral stability and maximal entropy principle}

The matrix ensembles discussed in this paper are
stable with respect to matrix addition in the sense
that the eigenvalue distribution for the 
matrix constructed as a sum of 
two independent matrices from the original ensemble 
$H=H_1+H_2$ is identical as the the original one up to 
a trivial rescaling.
Wigner-L\'evy matrices are obviously stable, since
the probability distribution for individual matrix elements
of the sum $H_{ij} = H_{1,ij}+H_{2,ij}$ is stable. 
A sum of two Wigner-L\'evy matrices is again a Wigner-L\'evy
matrix. The Wigner-L\'evy matrices are not rotatationally 
invariant. This means in particular 
that the eigenvalue distribution itself
does not provide the whole information about the underlying
matrix ensemble. Indeed, if $O$ is a fixed orthogonal matrix,
and $H$ is a Wigner-L\'evy matrix, then the matrix $OHO^T$
is not anymore a Wigner-L\'evy matrix but it has
exactly the same eigenvalue distribution as $H$.
In other words, an ensemble of Wigner matrices is not 
maximally random among ensembles with the same
eigenvalue distribution.
One expects that maximally random ensemble with the given
spectral properties should be rotationally invariant. In this
case one also expects that the stability holds
not only for the sum $H=H_1+H_2$ but also for the sum 
of relatively rotated matrices: 
$H= H_1 + O H_2 O^T$,
where $O$ is an arbitrary orthogonal matrix.
It can be shown \cite{BAL} that the ensemble of random matrices
which maximizes randomness (Shannon's entropy) for 
a given spectral density has the probability measure
exactly of the form (\ref{frmeasure}) as discussed here.

Stable laws are important because they define domains
of attractions. For example, if one thinks of a matrix addition 
one expects that a sum of many independent
identically distributed random matrices 
$H = H_1 + \dots + H_n$ should for $n \rightarrow \infty$ 
become a random matrix from a stable ensemble.

Maximally random spectrally stable ensembles which we discussed
in the section on free random matrices play a special role 
since they can serve as an attraction point for the sums of
iid rotationally invariant matrices. Moreover one expects 
that even for not rotationally invariant random matrices 
$H_i$, the sums of the form 
$B = O_1 H_1 O_1^T  + \dots + O_n H_n O_n^T$ where
$O_i$ are random orthogonal matrices, will for large $n$
generate a maximally random matrix $B$ from a spectrally
stable ensemble. In this spirit one can expect that
if ones adds many randomly rotated Wigner-L\'evy matrices:
\be
B = \frac{1}{{\cal N}^{1/\mu}}\sum_i^{\cal N} O_i A_i O_i^T
\ee
that for ${\cal N}\to \infty$ the matrices $B$ should 
become rotationally invariant, maximally random with
a distribution governed by the FRL symmetric distribution. 
In Figures 12-14 we show that this is indeed the case.
The plots illustrate the two types of stability discussed above. 
In each case we generate $N=100$ WL matrices 
and combine ${\cal N}=100$ of them either as a simple sum (black) or a 
rotated sum (red) with the appropriate scale factor. The plots
represent the numerically measured spectra for the two cases.
We present results for $\mu=1.5,~1.25$ and 1, which all show that a simple sum
reproduces the BC result, while the rotated sum reproduces the symmetric
FRL distribution.
\begin{figure}
\psfrag{L}{\bf{\Large $\lambda$}}
\psfrag{R}{\bf{\Large $\rho(\lambda)$}}
\centerline{\scalebox{0.6}{\rotatebox{0}{\includegraphics{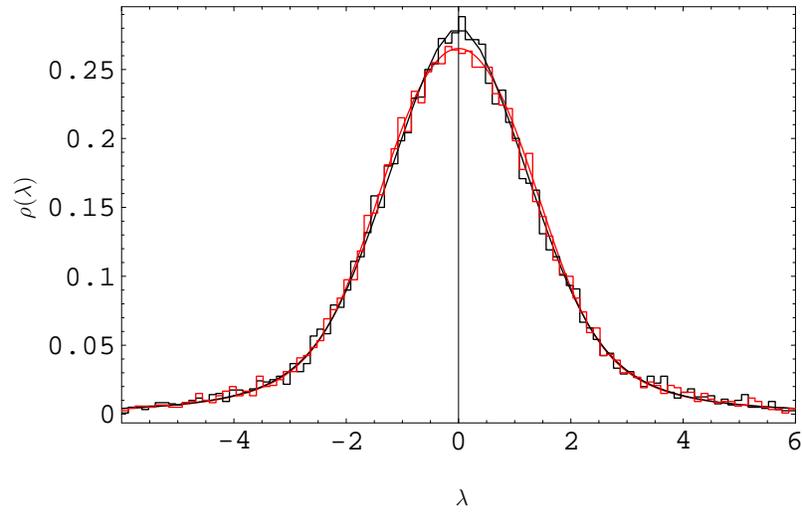}}}}
\caption[phased]{{\small WL (black) versus FRL (red) stability
for $\mu=1.5$}} \label{fig12}
\end{figure}
\begin{figure}
\psfrag{L}{\bf{\Large $\lambda$}}
\psfrag{R}{\bf{\Large $\rho(\lambda)$}}
\centerline{\scalebox{0.6}{\rotatebox{0}{\includegraphics{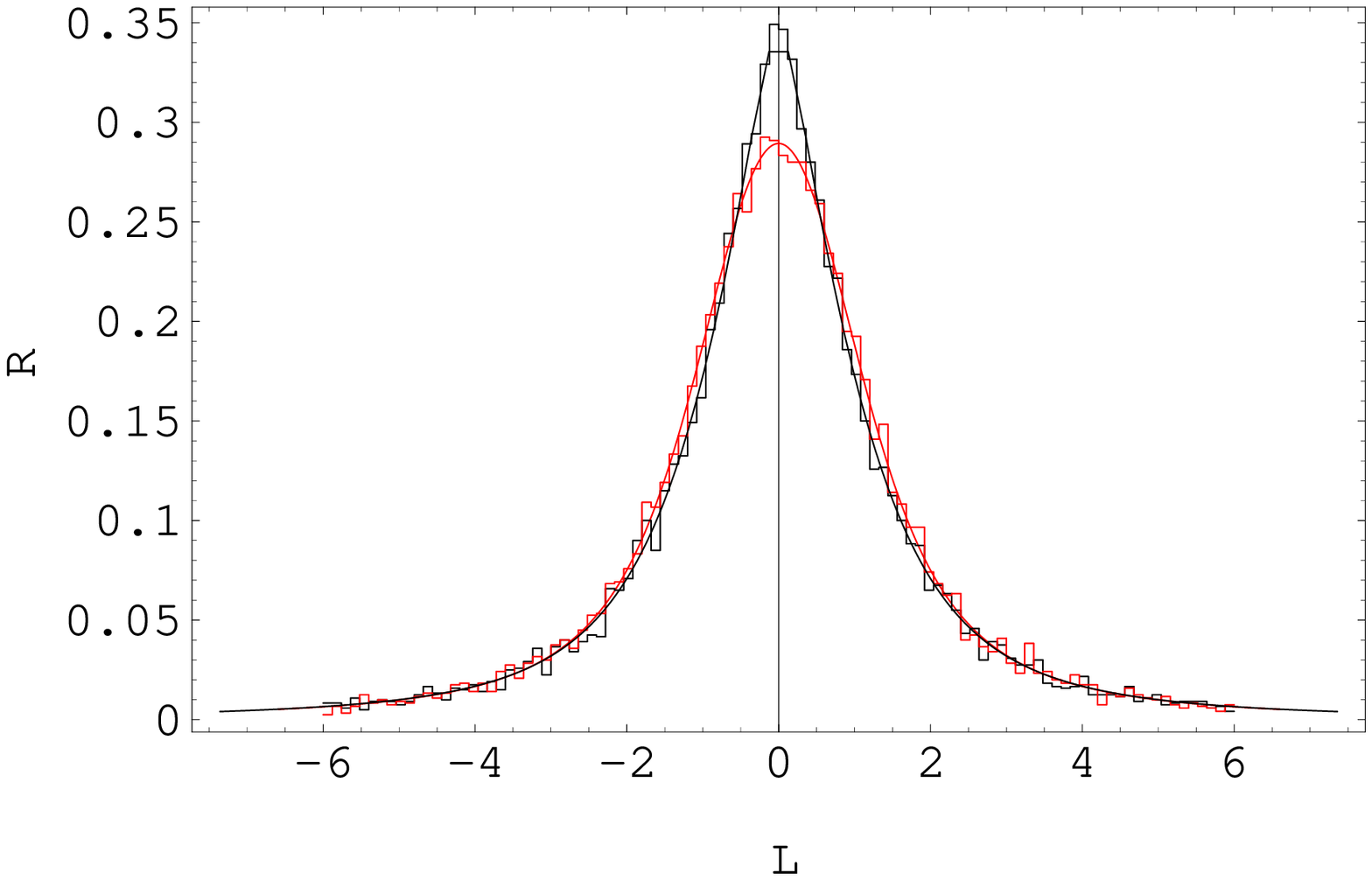}}}}
\caption[phased]{{\small WL (black) versus FRL (red) stability
for $\mu=1.25$}} \label{fig13}
\end{figure}
\begin{figure}
\psfrag{L}{\bf{\Large $\lambda$}}
\psfrag{R}{\bf{\Large $\rho(\lambda)$}}
\centerline{\scalebox{0.6}{\rotatebox{0}{\includegraphics{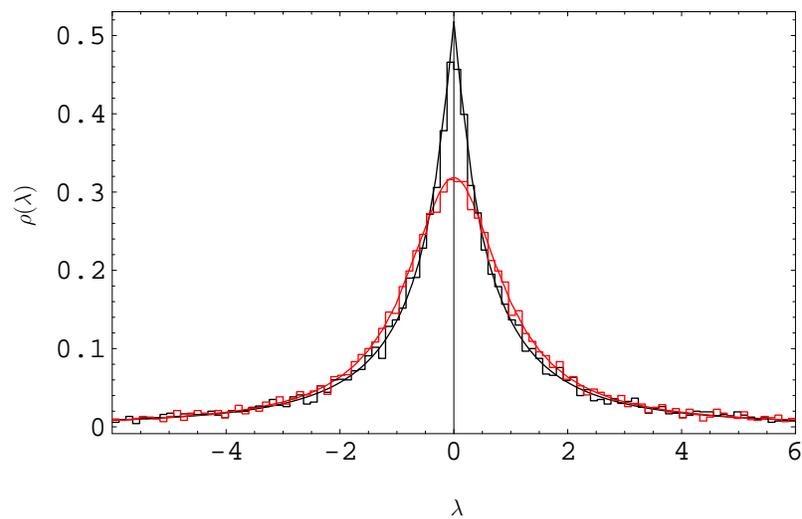}}}}
\caption[phased]{{\small WL (black) versus FRL (red) stability
for $\mu=1$}} \label{fig14}
\end{figure}
\section{Conclusions}

We have given a detailed analysis of the macroscopic limit of
two distinct random matrix theories based on L\'{e}vy type
ensembles. The first one was put forward by Bouchaud and Cizeau
\cite{BC} and uses a non-symmetric measure under the orthogonal
group, and the second one was suggested by us~\cite{FREELEVY} and
uses a symmetric measure.

After correcting the original analysis
in~\cite{BC}, in particular our formulae
(\ref{correct1}), (\ref{correct}) replace (10b) and (12b) in~\cite{BC},
and their eq. (15) is replaced by the pair of ``dispersion relations''
(\ref{G}) {\em and} (\ref{disp}), we  found perfect agreement between
the analytical and numerical spectra. The WL measure is easy to
implement numerically for arbitrary asymmetry parameter $\beta$
in the L\'{e}vy distributions. The spectrum of
WL matrices does not depend on $\beta$ and remains symmetric and universal,
depending only on $\mu$.

We have also shown that the spectra generated analytically for symmetric
FRL matrices are similar to the ones generated from WL matrices.
Unlike the WL ensemble, the FRL ensemble allows for
both symmetric and asymmetric L\'{e}vy distributions. Both ensembles
are equally useful for addressing issues of recent
interest~\cite{LEVYFIN,BPBOOK2}.

Let us finish the paper with two remarks. 
\begin{itemize}
\item {\bf i.} The application of free probability calculus to asymptotically free
matrix realizations allows one to derive spectral density of the
matrices from the underlying matrix ensemble
but it does not tell one how to calculate eigenvalue correlation
functions, or joint probabilities for many eigenvalues. Actually
different matrix realizations of free random variables may have
completely different structure of eigenvalue correlations even
if they are realizations of the same free random variables.
To fix correlations or joint probabilities for 
two or more eigenvalues, one has to introduce the concept
of higher order freeness \cite{2F}. We think however that 
if one imposes on a matrix
realization of free random variables an additional requirement
that it has to be maximally random in the sense of maximizing
Shannon's entropy \cite{BAL} then this aditional
requirement automatically fixes the probability
measure (\ref{frmeasure}) for the ensemble
and thus also all multi-eigenvalues correlations.

\item {\bf ii.} Large eigenvalues behave differently for Wigner-L\'evy and
maximally random free random L\'evy matrices
discussed in this paper. As pointed out recently \cite{SF},
the largest eigenvalues fluctuate independently 
for Wigner-L\'evy ensemble, a little bit like in the mean-field
argument \cite{BC} mentioned before, while for the maximally 
random matrix ensemble (\ref{frmeasure}) even large eigenvalues
are correlated \cite{FREELEVY}. 

\end{itemize}


\begin{acknowledgments}
We thank Jean-Philippe Bouchaud for discussions
which prompted us to revisit and compare the two approaches
presented in this paper. This work was partially supported
by the Polish State Committee for Scientific Research (KBN) grant
2P03B 08225 22 (MAN, ZB), Marie Curie TOK programme "COCOS -
Correlations in Complex Systems" (Contract MTKD-CT-2004-517186),
(MAN, JJ, ZB, GP), the National Office for Research and Technology
grant RET14/2005 (GP) and by US-DOE grants DE-FG02-88ER40388 and
DE-FG03-97ER4014 (IZ).
\end{acknowledgments}

\end{document}